# Physics-informed Neural-Network Software for Molecular Dynamics Applications

**Authors**

Taufeq Mohammed Razakh[a,b], Beibei Wang[a,c], Shane Jackson[a,c], Rajiv K. Kalia[a,b,c,d], Aiichiro Nakano[a,b,c,d,f], Ken-ichi Nomura[a,d], Priya Vashishta[a,b,c,d]

[a] *Collaboratory for Advanced Computing and Simulations, University of Southern California, Los Angeles, CA 90089-0241, USA*

[b] *Department of Computer Science, University of Southern California, Los Angeles, CA 90089-0241, USA*

[c] *Department of Physics & Astronomy, University of Southern California, Los Angeles, CA 90089-0241, USA*

[d] *Department of Materials Science and Chemical Engineering, University of Southern California, Los Angeles, CA 90089-0241, USA*

[f] *Department of Biological Science, University of Southern California, Los Angeles, CA 90089-0241, USA*

## Abstract

We have developed a novel differential equation solver software called PND based on the physics-informed neural network for molecular dynamics simulators. Based on automatic differentiation technique provided by Pytorch, our software allows users to flexibly implement equation of atom motions, initial and boundary conditions, and conservation laws as loss function to train the network. PND comes with a parallel molecular dynamics (MD) engine in order for users to examine and optimize loss function design, and different conservation laws and boundary conditions, and hyperparameters, thereby accelerate the PINN-based development for molecular applications.

## Keywords

*Molecular Dynamics; Machine Learning; Differential Equation Solver;*

## 1. Motivation and Significance

Molecular dynamics (MD) simulation is a vital tool in physics, chemistry, biomedical, and materials researches because it provides profound understanding with atomistic-level insights. While the size of simulation system and attainable temporal scale is usually in trade-off, trillion particles simulations have been used to simulate mechanical responses of materials using high-performance computing (HPC) cluster [1]. Also, numerous efforts have been made to extend the accessible temporal scale by atomistic simulation methods. Kinetic Monte Carlo [ref], hyperdynamics [ref], and parallel replica [ref] are well-known algorithms and successfully increased the accessible temporal scale by several orders of magnitude. However such time-accelerated algorithms often rely on the local characteristics of potential energy surface (PES) landscape, limiting its applicability to particular problems. A generic framework that extends accessible temporal scale has a great potential to enhance modeling capability in materials and biological simulations.

In MD simulation, Newton's equation of motions, a multi-dimensional coupled ordinary differential equation (ODE), is integrated by a numerical solver to obtain the trajectory of atoms. The maximum discretization unit in the time integral is dictated by physical properties of the target system, such as pressure, temperature, and phonon frequency. A serious bottleneck arises due to the sequential dependency in the time integration to solve the equations of atomic motion. Too large time step will lead the numerical solver unstable or drifts in conserved properties, resulting in the single time step no more than a few femtoseconds, i.e. $10^{-15}$ seconds. A variety of ODE solvers has been employed in MD simulations considering the accuracy, algorithmic robustness, and computational cost [refs], however, the parallelization in the time integral has not been addressed.

Due to recent rapid evolution and remarkable successes of Machine Learning (ML), Artificial Neural Network (ANN) has been attracting great attentions as a novel DE solver [5-9]. ANN as a differential equation (DE) solver, which was proposed by Dissanayake et al [9], encodes target DEs together with initial and boundary conditions as the loss function to minimize. Thanks to the universal approximation theory, feedforward NN is trained by following typical model training protocols to obtain the solution. This type of ANN is called physics-informed neural network (PINN) and has been successfully solved many DEs including heat equation [ref], Burger equation [ref], Schrodinger equation [ref], and Hamilton's equations of motions [ref]. Employing divide-and-conquer approach a recently proposed parallel PINN [10] combines a light-weight coarse-grained DE solver on top of smaller time segments that are solved concurrently to achieve greater time grid size with a negligible runtime increase due to model training. PINN holds a great promise for DE-based modeling for scientific and engineering applications, however, the degrees of freedom of systems has been demonstrated so far is relatively small, and its applicability to a practical MD simulation remain unclear due to the huge hyper-parameter space [11]. To this end, we have developed a portable, efficient and easy-to-use PINN-MD simulation software based on C++, Pytorch C++ Frontend, and Message Passing Interface (MPI) library. Our software comes with a scalable parallel MD engine, called pmd, allowing users to examine different model training scenarios, loss functions, and hyperparameters off-the-shelf, thereby enabling rapid development of PINN-MD algorithms.

## 2. Software Description

The heart of our software is developed with PyTorch C++ Frontend aiming to provide users high-performance, low latency, native multithreading support, the Fortran and C binding capability. These features are particularly suitable for many leadership-scale HPC software such as RXMD and QXMD to incorporate PND in their framework. Equipped with the automatic differentiation capability, PND allows uses to implement standard initial and boundary conditions, as well as many forms of physics-based constrain such as the conservation laws on the total energy and linear momentum, and the principle of least action [ref]. These constraints may be used to guide model training as heuristics to achieve a faster convergence of the loss function. To reduce the initial barrier for users to develop and integrate PND into their molecular simulation engine, we provide a scalable MD engine (pmd) and demonstrate a boundary condition problem solver with many constraints.

User defines the loss function to evaluate mean squared error (MSE) which is essentially the set of DEs governing the evolution of the system and constraints from boundary conditions. Solution of the DE is obtained by minimizing the MSE during ANN training, see Figure 1.

## 3.1 Directory Structure and Source Code Organization

In this section, we describe main classes and functions that support the core functionalities in PND. PND software is organized in three directories - ***MD_Engine***, ***Source*** and ***Example***. The header and source files to train neural network and solving the equations of motion are stored in ***Source*** directory. ***MD_Engine*** directory contains a scalable MD software (**pmd**) which may be replaced with user's own MD engine software. The class ***ScratchPad*** class inherits the **PND** class and expose member functions for users to customize the loss function. The ***ScrachPad*** source code is stored in ***Example*** directory to demonstrate techniques to perform MD simulation. An illustrative example using an FCC crystal MD simulation is presented in the Section 4.

Here we provide a list of the key classes, member functions and source codes with their brief explanation.

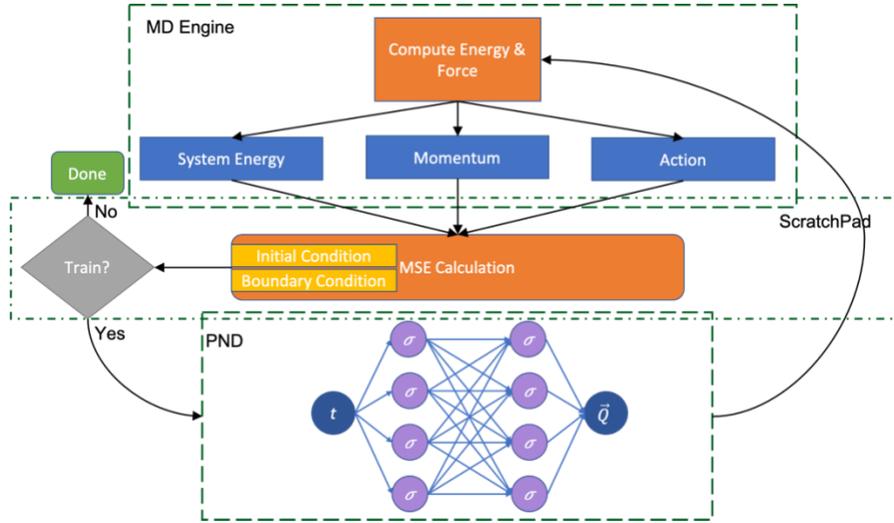

*Figure 1*: Schematic of PND workflow in molecular dynamics application. By inheriting the base class - PND into the users workspace (referred to as ScratchPad here) users can implement the laws for system in the form of a PDE through the interface of the superclass. The feed forward NN predicts atomic positions and velocities, which get passed to the MD engine to calculate terms which fit into the the systems PDE such as potential energy, total momentum. Implementing the PDE takes place by overriding the loss function method of the base class, thereby easily integrating the MD Engine layer and PINN training layer in their work space. The MSE calculation uses the sum of the mean-squared PDE residuals with automatic differentiation and the mean-squared error in initial and boundary conditions. Training of the PINN is carried out by minimizing the mismatch with respect to the NN parameters. For the given NN, t represents the time vector for which MD states need to be determined, σ the activation function and $\vec{Q}$ the predicted system states over t.

**Source/pnd.in:** The number of neurons in the neural network and training options such as epochs and learn rate are specified in this file.

**Source/pnd.cpp**: This source code defines functions which the user can use to structure and train a neural network for the purpose of solving DE's closely resembling an MD system over time steps. The functionality of the source code is encapsulated into the PND class. By instantiating this class users can access instance variables to store the training parameters. Instance methods are designed to update the instance's training parameters, thus only one instance of the class may be used for all training epochs. The PND class exposes virtual functions that implements the loss function. These functions are meant to overwritten by the user allowing flexible loss function designs depending on users' needs and simulation system.

**Source/pnd.hpp:** A header file with declarations for functions and variables of the class that gets defined in pnd.cpp.

**MD_Engine/pmd.cpp:** In this source code we define the class Atom and Subsystem. Collectively these classes provide functions and data structures to help with spatial decomposition and mapping the subsystem onto processors.

**MD_Engine/pmd.hpp:** A header file with declarations for classes that are that are defined in dependencied/pmd.cpp.

**MD_Engine/pmd.in:** The parameters for the MD system are defined in this file

**Example/ScratchPad.cpp:** This source code contains a sample implementation which interfaces with the class PND as well as the MD engine.

**CMakeLists.txt:** PND employs CMake build system. User needs to add the directories containing the MD and PND code as a search path for include files for our target.

## 3.2 Loss Function, Optimizer, and Training Driver Function

The derivative of the loss function with respect to the network input parameters are passed to the optimizer function ***UpdateParamsNadam*** before the weights and biases are adjusted to train the network. PND provides a predefined optimizer which implements the Nestrov and Adam algorithm (NADAM) implementation to train the NN. Advanced users may replace this function with various optimizer algorithms provided natively by PyTorch under the torch.optim class. Orchestrating this cycle of the loss calculation and network parameter optimization over training epochs is handled by the ***mainTrain*** function. The initial and boundary conditions as well as the systems energy from ground truth are passed down to this function as a tuple and consequently made available to the function calculating the MSE. We found that a pre-train step using a rough estimate of atom trajectory, such as linear extrapolation of positions and velocities would help the model training step. The use of the pre-training step also provides a better control and check to avoid unwanted atomic positions such as two atoms overlapping before entering into the main training step.

**Loss (params, icfs, energy)**
The loss function is computed in the function Loss, which takes the neural network parameters, initial and boundary conditions, and various constrains.

Parameters of the neural network, i.e., params, are trained under a loss function which represents a mismatch the solution vector based on the outputs of the NN show from Initial and boundary conditions, i.e., icfs, and also the systems total energy which is a conservative quantity. The returned value is the derivative of the error defined with respect to the params. For $d$-dimensional configuration space of $t$ time points, the initial position vector is represented by $\vec{q}_0$, the boundary position vector is represented by $\vec{q}_T$, the initial velocity vector is represented by $\vec{v}_0$ and the boundary velocity vector is represented by $\vec{v}_T$. For $Np$ atoms in a system, the system is represented with a vector of size $D = 2 \times d \times Np$. Outputs of the NN is represented as by $\vec{Q}(params) \in \mathbb{R}^{D \times t}$ for all time points and $\vec{Q}(params, t)$ output at any time $t$. The first half outputs of the NN $\vec{Q_1}(params)$ are vectors for the positions and the second half $\vec{Q_2}(params)$ are the velocities.

$$MSE^{\text{pre train}} = \left(\vec{Q_1}(params, 0) - \vec{q}_0\right)^2 + \left(\vec{Q_1}(params, T) - \vec{q}_T\right)^2 + \left(\vec{Q_2}(params, 0) - \vec{v}_0\right)^2 + \left(\vec{Q_2}(params, T) - \vec{v}_T\right)^2$$

$$MSE^{\text{main train}} = \left(\vec{Q_1}(params, 0) - \vec{q}_0\right)^2 + \left(\vec{Q_1}(params, T) - \vec{q}_T\right)^2 + \left(\vec{Q_2}(params, 0) - \vec{v}_0\right)^2 + \left(\vec{Q_2}(params, T) - \vec{v}_T\right)^2 + (predicted\ system\ energy - energy)^2$$

## 4. Illustrative Example

This section presents from results of training the neural network to closely resemble a simulated system whose parameters are the same as the sample shown in **Figure 2**. We evaluate the predicted systems by comparing the energies and atom trajectories of the simulated and predicted MD systems.

The simulation was carried out for Argon atoms which were brough to a relaxation state from an FCC orientation with density 0.8, initial temperature at 0.7(84K) time step of 0.01×2.2ps. The example demonstrates the predicted system strictly conserving the systems total energy while closely resembling the kinetic energy and potential energy over time steps with that of ground truth values (See Figure 2). Figure 3 presents a few typical atom trajectories that shows close resemblance to the ones obtained from MD simulation. An input file of this illustrative example is provided on the source repository within the Example directory.

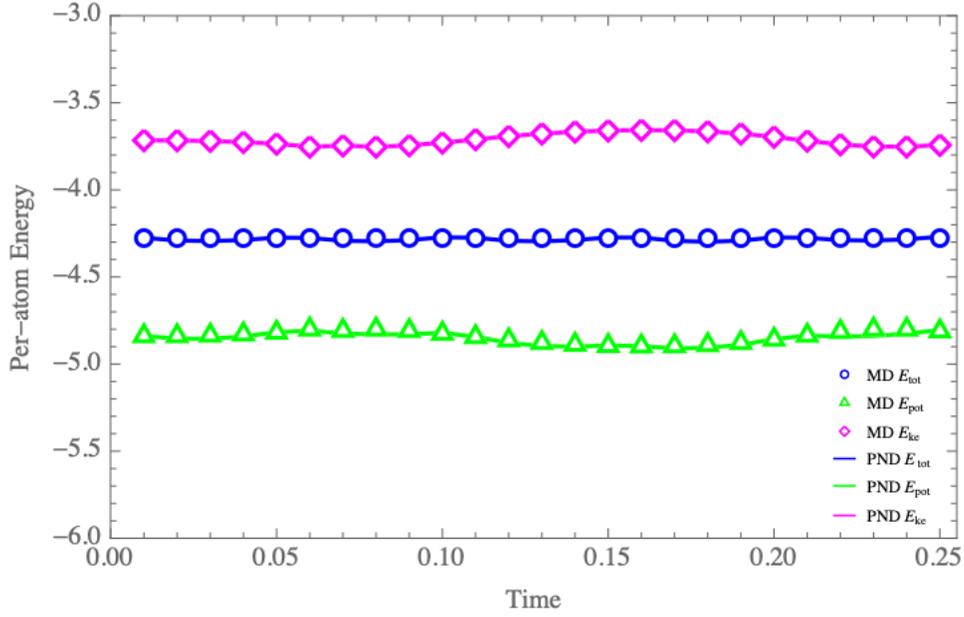

*Figure 2*: *Energy profile of PND (solid lines) and ground truth MD (markers) using a 32-atom FCC crystal with Lennard-Jones potential. The total, potential and shifted-kinetic energies of the system from prediction are shown in blue, green, and magenta respectively.*

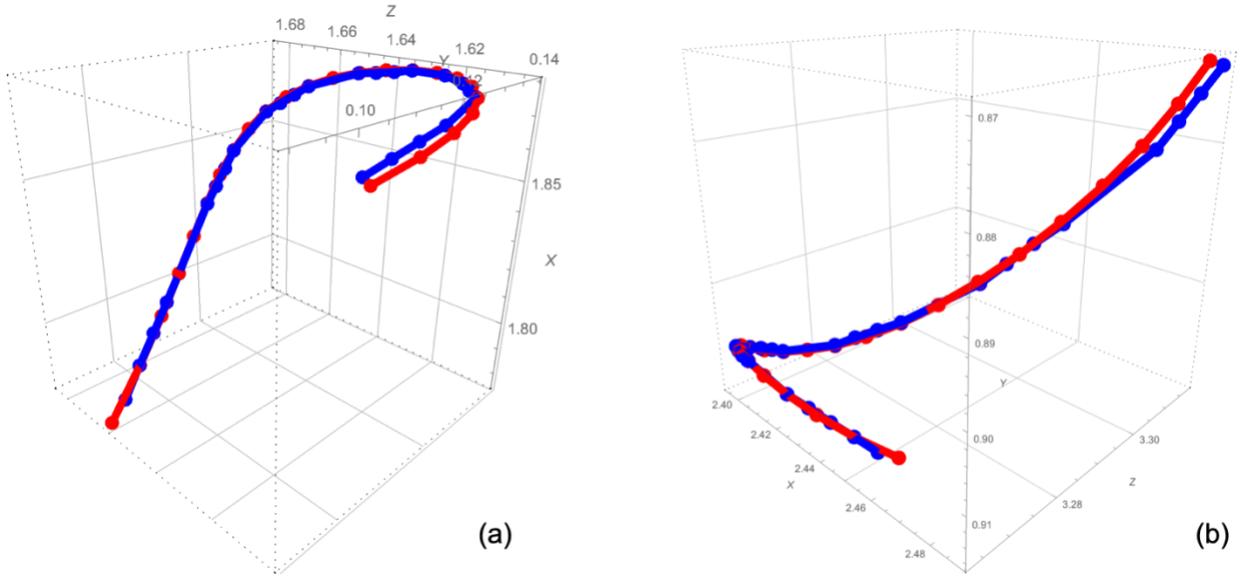

*Figure 3*(a) *and (b) are trajectories of randomly selected two atoms. Red trajectory corresponds to the ground truth MD simulations, blue trajectory is from PND.*

## 5. Impact

Neural network has been attracting as a novel and efficient DE solver and succesfully applied to many DE problems so far. PINN may address the long-standing problem in molecular simulation community, namely the sequential bottleneck from time integral. A number of studies has proven that the prediction accuracy and algorithmic concurrency may be further enhanced by exploiting the conservation laws specific in each simulation simulation. However the potential of the PINN approach has not been fully investigated because of the inevitable large degrees-of-freedom in molecular simulations, and more importantly, the lack of an off-the-shelf software that allows to simultaneously examine the effect of physic-informed loss function, model training performance, and assess the model prediction accuracy against the ground truth simulation results. Our work presented here aims to address the urgent demand and serves as a curucial step toward PINN-based molecular simulation methodlogy development.

## 6. Conclusions

Our open software allows researchers to continue with the application of physics informed neural networks (PINNs) to solve differential equations that govern MD systems. By using the interface provided by our code, users can train neural networks as well as gather and data for initial and boundary conditions using our lightweight MD code base. Our tool lays emphasis on experimenting with PDEs as loss functions and while limiting the interaction with MD code bases and scaffolding optimization code.

**Acknowledgements**
This work was supported as part of the Computational Materials Sciences Program funded by the U.S. Department of Energy, Office of Science, Basic Energy Sciences, under Award Number DE-SC0014607.

## B Required Metadata

### B1 Current executable software version

*Ancillary data table required for sub version of the executable software: (x.1, x.2 etc.) kindly replace examples in right column with the correct information about your executables, and leave the left columns as they are*

*Table 1 – Software metadata*

| Nr | (executable) Software metadata description | Please fill in this column |
|---|---|---|
| S1 | Current software version | 1.0 |
| S2 | Permanent link to executables of this version | https://github.com/TaufeqRazakh/PND |
| S3 | Legal Software License | GPL-3 |
| S4 | Computing platform / Operating System | Linux, OS X |
| S5 | Installation requirements & dependencies | GNU C++14, PyTorch 1.4.8 |
| S6 | Link to user manual | https://github.com/TaufeqRazakh/PND/blob/master/README.md |

| S6 | Support email for questions | razakh@usc.edu |

**B2 Current code version**

*Ancillary data table required for subversion of the codebase. Kindly replace examples in right column with the correct information about your current code, and leave the left columns as they are*

*Table 2 – Code metadata*

| Nr | Code metadata description | *Please fill in this column* |
|---|---|---|
| C1 | Current Code version | 1.0 |
| C2 | Permanent link to code / repository used of this code version | https://github.com/TaufeqRazakh/PND |
| C3 | Legal Code License | GPL3 |
| C4 | Code Versioning system used | Github |
| C5 | Software Code Language used | C++ |
| C6 | Compilation requirements, Operating environments & dependencies | Linux, OS X |
| C7 | If available Link to developer documentation / manual | https://github.com/TaufeqRazakh/PND/blob/master/README.md |
| C8 | Support email for questions | razakh@usc.edu |